\documentstyle{article}

\def\be{\begin{equation}}
\def\ee{\end{equation}}
\def\bea{\begin{eqnarray}}
\def\eea{\end{eqnarray}}
\def\ba{\begin{array}}
\def\ea{\end{array}}
\def\Del{\bigtriangledown}
\def\D2{\Del^2 \phi}
\def\G2{(\Del \phi)^2}
\begin{document}
\parindent 0pt
{\large A note on the three dimensional sine--Gordon equation}
\\ Ahmad Shariati
\\ Institute for Advanced Studies in Basic Sciences, P.O. Box 159
Zanjan 45195, Iran.
\\ Institute for Studies in Theoretical Physics and Mathematics, P.O. Box
5531 Tehran 19395, Iran.
\begin{abstract}
Using a simple ansatz for the solutions of the three dimensional
generalization of the
sine--Gordon and Toda model introduced by Konopelchenko and Rogers,
a class of solutions is found by elementary methods.
It is also shown that these equations are not evolution equations in the
sense that sotution to the initial value problem is not unique.
\end{abstract}
\parskip 10pt
Integrable models in more than two independent variables are of interest
\cite{AC}. One class of such models is the generalized Loewner systems
introduced by Konopelchenko and Rogers \cite{KR1,KR2}. These are third
order partial differential equations which are first order in $z$ which
is to be the temporal variable. These equations are studied using
sophisticated methods such as Lie--B\"acklund transformation \cite{KSR},
$\bar{\partial}$ dressing method \cite{KR2,DK}, and Painlev\'e analysis
\cite{RL}.

In the following we consider the three dimensional generalizations
of the sine--Gordon and Toda model as introduced in \cite{KR2}.
The observation is that a large class of
solutions to these partial differential equations and an important property
of these equations can be obtained by a very simple and elementary
method.

We begin with the so called three dimensional sine--Gordon equation \cite{KR2} which
reads as
\be \label{1}
   \left( \frac{\theta_{zx}}{\sin \theta} \right)_x
 - \left( \frac{\theta_{zy}}{\sin \theta} \right)_y
 + \frac{\theta_x \theta_{zy} - \theta_y \theta_{zx}}{\sin^2 \theta}
 = 0 \ee
 where a subscript means partial differentiation. We use the following
 ansatz.
\be \label{2} \theta(x,y,z) = f(u(x,y),z). \ee
 This makes the last term in \ref{1} identically vanish and equation
 \ref{1} becomes
\be \label{3}
   (u_{xx} - u_{yy}) \frac{f_{uz}}{\sin f}
 + (u_x^2 - u_y^2 ) \left( \frac{f_{uz}}{\sin f} \right)_u = 0.
\ee
 From this it follows that equation \ref{1} has two solutions as follows:
 \\  A: $\theta(x,y,z) = f(u(x,y),z)$ for $u(x,y) = g(x+y) +
 h(x-y)$, where $g$ and $h$ are two arbitrary functions and
 $f$ is a solution to the two dimensional sine--Gordon eqaution $f_{uz} = c \sin f$
 in lightcone coordinates. Here $c$ is a constant.
 \\ B: $\theta(x,y,z) = f(x \pm y, z)$ for any arbitrary function
 $f$ of two variabales.

 We use these results to get some information about this partial
 differential equation.
 Consider a function $h(z)$ which is smooth and has the following
 properties.
$ h(z) = 0$ for $z \le z_0$  where $z_0 > 0$; and
$h(z) \ne 0$ for $z > z_0$.
Now, the function
$\theta(x,y,z) = f(x\pm y) h(z)$ for an arbitrary $f$ has the following
properties: 1) $\theta$ is a solution to \ref{1}, and 2) $\theta(x,y,z) =0$
for $z \le 0$. This shows that the solution to the initial value problem
for equation \ref{1} is not unique.

This non-- uniqnueness makes it
unpleasant as an evolution equation. This is not surprising, for $z$ is
not `time'. $z$ is something like one of the light cone (characteristic)
variables. To get an idea of what is happening consider the wave equation
in light cone coordinates.
The equation $\theta_{uz}=0$ has the known solution
$\theta(u,z) = f(u) + g(z)$, where $f$ and $g$ are two arbitrary
functions. Now the `initial' value problem $\theta(u,0) = f(u)$ has
infinite solutions: Any function $g(z)$ such that $g(0)=0$ gives
a solution $\theta(u,z) = f(u) + g(z)$. In other words, specifying the
value of $\theta$ on a lightlike line does not lead to a unique solution;
one has to specify the value of $\theta$, considered as a fuction of
$(t,x)=(u-z,u+z)$,
and its first temporal derivative on a spacelike line to get a unique
solution. Therefore, the
argument preceding this paragraph shows that equation \ref{1} is not a
good evolution equation in the sense that $z$ is not a temporal variable.

The non--uniqueness of the solution to the initial value problem
of equation \ref{1} can also be
seen from type A solutions as follows. The sine--Gordon equation
$f_{uz} = \sin f$ has the following solution
\be \label{S} f(u,z;\alpha) = 4 \tan^{-1}
e^{ \alpha^{-1} u +\alpha z } \ee
where $\alpha$ is a non--zero real parameter.
Take a function $u(x,y)$ which is a solution to the wave equation and
construct the following two solutions to equation \ref{1}
\be
\theta(x,y,z) = f(\alpha u(x,y),z;\alpha) =
4 \tan^{-1} e^{u(x,y) +\alpha z} \ee
\be
\theta(x,y,z) = f(u(x,y),z;1) =4 \tan^{-1} e^{u(x,y) + z} \ee
These two solutions agree at $z=0$ but disagree at $z \ne 0$.

Now we turn to the three dimensional generalization of the Toda model
which is given by a $2 \times 2$ Cartan matrix $K^{\alpha \beta}$.
The same ansatz leads to solutions for this model.
The differential equatin is
\be \label{5}
\left( e^{-\sum_{\beta}K^{\alpha \beta}\psi_{\beta}}\psi_{\alpha zx}
\right)_x - \sigma^2
\left( e^{-\sum_{\beta}K^{\alpha \beta}\psi_{\beta}}\psi_{\alpha zy}
\right)_y = 0 \quad \quad \alpha = 1, 2. \ee
Here $\alpha$ is an index while subscripts $x$, $y$, and $z$
denote partial differentiation as usual  . Now we use the following ansatz:
\be \psi_{\alpha}(x,y,z) = f_{\alpha}( u(x,y), z). \ee
Equation \ref{5}, then, becomes
\be \label{6}
(u_{xx} - \sigma^2 u_{yy}) e^{-\sum_{\beta}K^{\alpha \beta}f_{\beta}}
+ (u_x^2 - \sigma^2 u_y^2)
\left( e^{-\sum_{\beta}K^{\alpha \beta}f_{\beta}}f_{\alpha zu} \right)_u
= 0. \ee
From this it is evident that equation \ref{5} has the following two
solutions: \\ A: $\psi_{\alpha}(x,y,z) = f_{\alpha}(u(x,y),z)$
where $u(x,y) =g( x +\sigma^{-1} y) + h( x - \sigma^{-1} y)$ for two
arbitrary functions $g$ and $h$ provided that $f_{\alpha}(u,z)$ are
solutions to the two dimensional Toda equation
$f_{\alpha zu} = c e^{-\sum_{\beta}K^{\alpha \beta}f_{\beta}}$. Here
$c$ is a constant. \\  B:
$\psi_{\alpha}(x,y,z) = f_{\alpha}( x \pm \sigma^{-1}y, z)$ for arbitrary
functions $f_{\alpha}$ of two variables.

We conclude this letter by the following remarks.
\\ 1. On setting $\sigma =0$ in \ref{5} we get a third order partial
differential equation in two variables $x$ and $z$, which after a first
integration leads to
$\psi_{\alpha z x}=c(z) e^{\sum_{\beta}K^{\alpha \beta}} \psi_{\beta}$,
 where $c$ is an arbitrary function of $z$. By redefining $z$,
$c$ can be considered as a constant. This is the familiar two dimensional
Toda model. Therefore, it must be true that the $\sigma \to 0$ limit of
solutions A and B lead to solutions of the two dimensional model. In doing
so one must be careful because the wave equation $u_{xx} -\sigma^2 u_{yy}=0$
in that limit leads to $u=a x + c$, for two constants $a$ and $c$. In fact,
$c$ may be a function of $y$ but this dependence is not relevant.
Therefore, type A solutions actually become the solutions of the two
dimensional model. For type B solutions, we note that the limit of
$u_x^2 -\sigma^2 u_y^2 =0$ as $\sigma \to 0$ leads to $u=$constant.
Now, a glance at 8 shows that type B solutions, in this limit,
read as $\psi_{\alpha}(x,y,z)=f_{\alpha}(z)$ for arbitrary $f_{\alpha}$
which is trivially a solution of 7. This shows that type B solutions
are peculiar to the third order equation 7.
\\ 2. Setting $\sigma^2 \to -\sigma^2$ in \ref{5} leads to a three dimensional
model which has solutions of type A for  any harmonic function $u(x,y)$.
Solutions of type B disappear. The same is true for the three dimensional
sine--Gordon equation: \\
On changing the sign of the second term in equation \ref{1},
one gets the following partial differential
equation which has solutions of type A
if $u(x,y)$ is a harmonic function and $f$ is a solution of the two
dimensional sine--Gordon equation.
\be
   \left( \frac{\theta_{zx}}{\sin \theta} \right)_x
 + \left( \frac{\theta_{zy}}{\sin \theta} \right)_y
 + \frac{\theta_x \theta_{zy} - \theta_y \theta_{zx}}{\sin^2 \theta}
 = 0 \ee
This partial differential equation has SO(2) symmetry
in the $(x,y)$ variables while equation \ref{1} has SO(1,1) symmetry, as
is shown in \cite{KR2}. The non--uniqueness of the initial value problem
remains intact, therefore, these Euclidean versions are not
evolution equations either.

Acknowledgments \\ My thanks go to A. Aghamohammadi, M. Korrami,
M. Alimohammadi, and B. Farnoudi
for fruitful discussions.

\end{document}